\documentclass[11pt]{iopart}
\usepackage[dvips]{graphicx}

\usepackage{amssymb}
\usepackage{amsthm}
\usepackage{amscd}
\usepackage{amstext}
\usepackage{amsfonts}
\usepackage{hyperref}
\newcommand{\p}{\partial}

\newcommand{\dd}{{\rm d}}

\newtheorem{theorem}{Theorem}[section]

\newtheorem{remark}[theorem]{Remark}

\begin{document}

\title{{ Weak~gauge~principle~and~electric~charge~quantization }} 
%
\vspace{0.5cm}
\author{E. Minguzzi$^{1,2}$\footnote{New permanent address: Department of Applied Mathematics, Florence University
Via S. Marta 3, 50139 Florence, Italy.\\ Work published by Journal
of Physics A: Mathematical and General $\copyright$ (2006) IOP
Publishing Ltd. http://www.iop.org.}, C. Tejero Prieto$^1$ and A.
L\'opez Almorox$^1$ }

\address{$^1$ Departamento de Matem\'aticas,
Universidad de Salamanca,  Plaza de la Merced 1-4, E-37008 -
Salamanca, Spain}
\address{$^2$ INFN, Piazza dei Caprettari 70, I-00186
Roma, Italy} \eads{\mailto{minguzzi@usal.es},
\mailto{carlost@usal.es}, \mailto{alm@usal.es}}

%
%
%
%


\begin{abstract} \\
Starting from a weak gauge principle we give a new and critical
revision of the argument leading to charge quantization on arbitrary
spacetimes. The main differences of our approach with respect to
previous works appear on spacetimes with non trivial torsion
elements on its second integral cohomology group. We show that in
these spacetimes there can be topologically non-trivial
configurations of charged fields which do not imply charge
quantization. However, the existence of a non-exact electromagnetic
field always implies the quantization of charges. Another
consequence of the theory for spacetimes with torsion is the fact
that it gives rise to two natural quantization units that could be
identified with the electric quantization unit (realized inside the
quarks) and with the electron charge. In this framework the color
charge can have a topological origin, with the number of colors
being related to the order of the torsion subgroup. Finally, we
discuss the possibility that the quantization of charge may be due
to a weak non-exact component of the electromagnetic field extended
over cosmological scales.
\end{abstract}

\maketitle


\section{Introduction}

The idea of justifying electric charge quantization by means of
topological arguments goes back to Dirac \cite{dirac31}. He showed
that the existence of a magnetic monopole would imply that both
the charge of the monopole and the electric charge of any other
particle in the Universe are quantized. Since the electromagnetic
field diverges at the worldline of the magnetic monopole, the
problem was essentially that of a charged particle moving  on a
spacetime with non-trivial topology.

Wu and Yang presented in \cite{wu75} a geometrical explanation for
the charge quantization argument proposed by Dirac. These authors
showed that the quantization of the charge of a magnetic monopole
is completely equivalent to saying that the monopole can be
described as a connection on a non trivial principal $U(1)$-bundle
over spacetime. Although the approach followed by these authors is
suitable for the particular case of the magnetic monopole, as we
shall see in due course their point of view is not general enough
for studying charge quantization on arbitrary spacetimes.

Therefore, the purpose of our paper is to analyze, starting from
basic and well established physical principles, the most general
conditions which lead to quantization of electric charge on
spacetimes with non-trivial topology. According to recent research
these spaces can not be directly ruled out if we take into account
our present cosmological knowledge
\cite{lachiezerey95,levin98,luminet03}.

We give a new and critical revision of the argument leading to
charge quantization based on what we call the ``weak gauge
principle'' and without assuming a priori an identification between
electromagnetism and principal $U(1)$-bundles with connections. The
main difference of our approach with respect to the standard ones
lies on the fact that the weak gauge principle only appears in the
presence of (at least) two different charged matter fields in
interaction with a gauge field. Notice that one of them should be
considered as the charged reference field allowing for a measure of
the relative interference Aharonov-Bohm class with respect to the
second charged field.

As we shall see, this extension of the gauge principle agrees with
the ordinary gauge principle (referred to in the paper as the
``strong gauge principle'') and makes no difference for spacetimes
with trivial topology. In this class of spaces leads to the same
results already present in the literature. However, the weak gauge
principle has non trivial physical implications on spacetimes with
non-trivial second integer cohomology group with non-vanishing
torsion. Notice that this is a radical difference with previous
works on this matter which have been based on manifolds without
torsion.

Now let us briefly explain some of the arguments that have lead us
to introduce the weak gauge principle. On Minkowski spacetime, a
charged matter field $\psi$ with charge $q_\psi$ changes under a
gauge transformation $A^\prime =A+d\alpha$ as
$\psi^\prime=e^{iq_\psi\alpha}\psi$. If we consider another charged
matter field $\phi$ with charge $q_\phi$, then under the same gauge
transform it changes as $\phi^\prime=e^{iq_\phi\bar\alpha}\phi$ with
$d\alpha=d\bar\alpha$. Since the spacetime is contractible we have
$\bar\alpha=\alpha+\mathrm{cnst.}$ and the difference in the gauge
transformation of the two fields is a global phase which can be
gauged away after a redefinition of $\phi$.  Therefore, in this
spacetime the usual formulation of the $U(1)$ gauge invariance of
the theory for two different charged matter fields  is simply
written as
\[
{A'}=A+\dd \alpha, \qquad {\psi'}= e^{i q_{\psi} \alpha} \psi,
\qquad {\phi'}=e^{i q_{\phi} {\alpha}} \phi , \] with the same
function $\alpha$ for both fields (the strong gauge principle).

On the other hand, it is well known that on a general spacetime
$M$ the gauge field $\{A^i\}_{i\in I}$ and the charged matter
fields $\{\phi^i\}_{i\in I}$, $\{\psi^i\}_{i\in I}$ are described
by families parametrized by a good covering $\{U_i\}_{i\in I}$ of
$M$ (see Section \ref{gau-prin-ntt} for more details).
Therefore the transformation of the fields under a gauge
transformation on $U_{ij}=U_i \cap U_j$ between $U_j$ and $U_i$
should be written as
\[
{A}^{i}=A^{j}+\dd \beta^{ij}, \qquad {\psi}^{i}= e^{i q_{\psi}
\beta^{ij}} \psi^{j}, \qquad {\phi}^{i}=e^{i q_{\phi}
{\bar\beta}^{ij}} \phi^{j}. \] Although for each contractible open
set $U_{ij}$ we still have
$\beta^{ij}=\bar\beta^{ij}+\mathrm{cnst.}$, notice that in general,
a non-trivial topology of the spacetime might make impossible to
globally gauge away the difference between the constants
$\{\beta^{ij}\}_{i\in I}$, $\{\bar\beta^{ij}\}_{i\in I}$. Indeed a
$U(1)$ redefinition of $\phi^{j}$ affects $\bar\beta^{ij}$ not only
for a given $i$ but for every value of the index $i$. In these
conditions we say that the two families of fields are related by a
``weak gauge transformation''. These observations naturally lead us
to introduce the notion of {\em weak gauge equivalence} for the
possible configurations $\{A^i,\phi^i,\psi^i\}_{i\in I}$, of the
system formed by the gauge field and the charged matter fields.

Let us point out that the weak gauge principle could also be hinted
from a study of the symmetries of the Lagrangian of the Standard
Model of particle physics. In this theory, matter fields interact
through the mediation of vector bosons, as a consequence the
Lagrangian is invariant under weak gauge transformations, that is,
there is  a global $U(1)$ symmetry associated to each matter field.
On the contrary  in a generic Lagrangian, with mixed neutral
interaction terms as, say, $\psi^{3}(\phi^*)^{2}$, with $3
q_{\psi}-2q_{\phi}=0$, there is only a global $U(1)$ symmetry that
involves all the matter fields at the same time, that is
$\alpha=\bar{\alpha}$. The weak gauge principle naturally embodies
the $U(1)$ matter field dependent redefinition freedom of the
Standard Model, whereas the strong gauge principle
does not. 

In this context the weak gauge principle establishes the
``intersection rule'' that expresses the relationship between
fields defined on any two overlapping open sets of the covering.
From a mathematical point of view the intersection rule says that
the charged matter fields are sections of Hermitian complex line
bundles (or equivalently principal $U(1)$-bundles). However, let
us point out that the gauge field $\{A^i \}_{i\in I}$ does not
determine, in general, a principal $U(1)$-bundle. This is due to
the following two reasons:

\begin{itemize}
\item[a)] On the first place, the bundle may not exist at all.
The family of 2-forms ${\{dA^i\}}_{i \in I}$ determines a global
closed 2-form $F$ on $M$. It is well known that one can only
associate with it a $U(1)$-principal bundle under appropriate
integrality conditions.

\item[b)] Secondly, even if the principal $U(1)$-bundle exists,
it might be not unique, since it is only determined up to flat
bundles on $M$. It is well known that the family of isomorphism
classes of flat $U(1)$-bundles is parametrized by the cohomology
group $H^1(M,U(1))$. In particular the non-trivial flat bundles
are in one-to-one correspondence with the torsion elements of
$H^2(M,\mathbb Z)$.
\end{itemize}

But even for the cases in which a principal bundle does exist, the
weak gauge principle requires to consider the whole family of
principal $U(1)$-bundles determined by the gauge field. In fact,
given a pair of charged fields $\psi$, $\phi$, each of them
determines a line bundle $P_\psi$, $P_\phi$. Now the weak gauge
principle is equivalent, in this case, to saying that the line
bundle $P_\psi$ differs from $P_\phi$ by a flat line bundle $K$.
That is $P_\psi=K\otimes P_\phi$, where $K$ represents, in
geometrical terms, the holonomy difference of these two line
bundles.

From a physical point of view, the class $[K]\in H^{1}(M,U(1))$
measures the relative Aharonov-Bohm interference of $\psi$ and
$\phi$, which controls the different behavior of the particles under
topological Aharonov-Bohm experiments. If this class is trivial the
phenomenology reduces to  the standard one given by the strong gauge
principle.

As a consequence, the line bundles $P_\psi$, $P_\phi$ which
describe the charged fields are in general not associated to the
same $U(1)$-bundle but to different members of the family of
$U(1)$-bundles determined by the gauge field. If $K$ is not
trivial, this in sharp contrast with the standard description
which assumes that all the charged fields are associated to the
same principal bundle. This is essentially the mathematical
content of the weak gauge principle. Therefore, the weak gauge
principle differs in an essential way from the ordinary gauge
principle in spacetimes $M$ with a non vanishing torsion subgroup
of $H^2(M,\mathbb Z)$.  To our best knowledge this fact has not
been previously recognized in the literature.

In the paper we analyze in detail the implications that the weak
gauge principle has for charge quantization. In order to make the
paper accessible to a wider audience, we have chosen to carry out
this analysis by means of \v{C}ech cocycles rather than using the
geometrical theory of bundles. We start just from the beginning
with a good covering so that the relevance of triple intersections
and \v{C}ech cohomology becomes clear. In some sense we continue
the generalization of Wu and Yang's paper made by Horv\'athy
\cite{horvathy84} who studied the monopole in a generic covering
by considering  the charge quantization argument in a generic
field configuration. Since we show explicitly how the involved
bundles  are defined, the physically oriented reader may find this
approach particularly clarifying while the mathematically oriented
reader may find interesting how the cocycle condition arises from
physical considerations based on the gauge principle and the
existence of a matter field.

We shall see that the simple existence of a non-exact
electromagnetic field on spacetime implies the quantization of
charges and therefore  the magnetic monopole is  just one of the
many possible non-exact field configurations leading to charge
quantization. This fact should not come as a surprise since it has
been recognized a number of times since Wu and Yang's work (see
also \cite{greub75,friedman79,balachandran80}) that the
quantization of the electric charge is related to the
classification of principal $U(1)$-bundles  in terms of Chern
classes.

However, a main difference of this paper with respect to previous
works is the fact that on spacetimes with torsion in the second
integral cohomology group, the non triviality of the bundles
associated to the charged fields does not necessary imply charge
quantization.

Another consequence of the theory for spacetimes with torsion is the
fact that it gives rise to two natural quantization units that we
identify with the electric quantization unit (realized inside the
quarks) and with minus the electron charge. In this framework the
color charge can have a topological origin, with the number of
colors being related to the order of the torsion subgroup of
$H^{2}(M,\mathbb{Z})$.

We also point out that the quantization of charge may be due to a
weak non-exact component of the electromagnetic field extended
over cosmological scales if at those scales a non-trivial topology
of the spacetime manifold arises. This component could have formed
in the initial instants of the Universe when its topology acquired
a final form. Then the expansion of the Universe would have
decreased its magnitude making it  undetectable in today
experiments.

Let us mention that recent discussions \cite{ghosh03,chatterjee04}
on the role of coverings in the deduction of Dirac's quantization
condition led us to believe that a treatment of charge
quantization in general coverings and for general spacetimes like
ours could help to clarify the assumptions that stay at the heart
of topological quantization.

In order to finish this introduction, we recall that over the
years other approaches have been introduced in order to give an
explanation of the quantization of the electric charge. They
involve topological arguments
\cite{ranada92,ranada98,kassandrov04,cohen02}, geometric
quantization \cite{sniatycki74,mitchell04}, path integral
considerations \cite{alvarez85}, anomaly cancellations
\cite{babu89,foot90,pires02,foot94,bowes96,foot99}, Kaluza-Klein
theory \cite{klein26b}, a particular analysis of the Aharonov-Bohm
potentials \cite{barone04}, loop quantization \cite{corichi98} or
a particular quantum theory of the electric charge
\cite{staruszkiewicz89}. The Dirac quantization condition can be
derived from the quantization of the total angular momentum
\cite{goddard78} and can be related to the associativity of finite
translations \cite{jackiw85,nesterov04}. However, there have been
also claims of inconsistency of Dirac's quantization condition in
second quantization \cite{he95}. Moreover, Schwinger suggested
that the magnetic charge should be an {\em even} integer of the
Dirac unit \cite{schwinger68,schwinger69}. These arguments are,
however, not generally accepted
\cite{poulis95,poulis95b,goddard78}.

\section{The gauge principle in Minkowski spacetime}
The theory of connections was developed by mathematicians in the
fifties and only in the seventies the relation with physics and
with the gauge principle was fully realized. Nowadays the
electromagnetic field is described by a connection on a principal
$U(1)$-bundle and a charged particle field is regarded as a global
section of an associated bundle which is constructed from a
representation of $U(1)$ on the vector space $\mathbb{C}$. Only
particles having a quantized charge can be implemented in this
mathematical setting and, for this reason, to recast the Dirac
original argument in modern terms means to justify why this
mathematical description is essentially the only one available.
Therefore, for the moment we  forget about the mathematics of
principal fibre bundles and connections and consider the simple
case of Minkowski spacetime. We define the gauge principle and
only later we move on to see what happens in  spacetimes with
non-trivial topology.

 Let $M$ be a spacetime with a metric $(+ - - -)$ and the topology of $\mathbb{R}^{4}$ (Minkowski
spacetime). The electromagnetic field $F$ is a closed 2-form field
on $M$. The matter field of a particle is  a mapping $\psi: M \to
\mathbb{C}$. We shall denote matter fields with the letters $\psi$
and $\phi$. We shall explicitly use two matter fields since the
quantization of charge is a relation between the charges of two
fields. The meaning  of what follows would be therefore clearer
working with two matter fields at the same time. Moreover, we
shall need at least two matter fields in order to distinguish
between the weak and the strong gauge principles (see below). It
is understood that  our study can be straightforwardly generalized
to any number of fields.

Since, the spacetime is contractible there is a potential (1-form)
field $A$ such that $F= \dd A$, moreover let $A'$ be another
potential, we have $\dd (A-A')=0$ and since $M$ is contractible
($M$ simply connected would suffice here) we have $A'=A+\dd
\alpha$, where $\alpha : M \to \mathbb{R}$ is a function.

By definition the triplets $(A, \psi, \phi)$,  $(A', \psi',
\phi')$, are related by a {\em weak gauge transformation} if there
exist functions $\alpha$, $\bar{\alpha}$ differing by a constant
$h=\alpha-\bar{\alpha}$ and such that
\begin{eqnarray}
A'&=&A+\dd \alpha , \label{oi}\\
\psi'&=& e^{i q_{\psi} \alpha} \psi ,\\
\phi'&=&e^{i q_{\phi} \bar{\alpha}} \phi .
\end{eqnarray}
If $h=0$ the two triplets are related by a {\em strong gauge
transformation}. The constant $q_{\psi}$ (resp. $q_{\phi}$) is the
electric charge of the matter field $\psi$ (resp. $\phi$). The
charges are quantized if there exists $q \in \mathbb{R}^+$ such
that $q_{\psi}=m q$, $q_{\phi}=n q$ with $n , m \in \mathbb{Z}$
coprime. Note that if $q$ exists then it is defined univocally by
the requirement that $m$ and $n$ are coprime (i.e. there are
integers $M$, $N$ such that $Mm+Nn=1$). Note also that by {\em
charges} we shall always mean those parameters that appear in the
gauge transformation. In general the Lagrangian depends on them
and therefore they will have some experimental consequences from
which their values can be recovered.  We stress that we do not
define the {\em charge} as the integral over a suitable surface of
the $0$-component of a certain conserved current. We also stress
that the word {\em quantization} is used as a synonym for {\em
discretization}. Making this choice we have followed the most
used, although sometimes misguiding, terminology.

Notice that in most treatments only one field is considered and
therefore the difference between weak and strong gauge
transformations can not be appreciated. Note also that the
definition of weak gauge transformation is symmetric between
$\psi$ and $\phi$ since $\alpha$ and $\bar{\alpha}$ differ by a
constant and therefore ${\alpha}$ can be replaced by $\bar\alpha$
in Eq. (\ref{oi}). In the presence of many fields one would need a
different function $\alpha$ for each field. In the following the
calculations will make sense in both the weak and strong cases.

There is no universally accepted definition of gauge principle but
the following definition seems to summarize the most relevant
features. A physical theory of the electromagnetic field satisfies
the gauge principle if \\

{\em {\bf (Weak/strong) Gauge principle}. The physical states of
the theory are the equivalence classes $[A,\psi,\phi]$ where two
elements $(A', \psi',\phi')$, $(A, \psi,\phi)$ are equivalent, $(
A', \psi',\phi')\sim( A, \psi,\phi)$ ,  iff they are related by a
(weak, strong) gauge transformation for a suitable function(s)
$\alpha$. The set of  physical states will be denoted $\{
[A,\psi,\phi]\}$.}\\


The difference between weak and strong gauge principles is not
important in the Minkowskian case since the Lagrangian is usually
invariant under symmetry transformations  that multiply a field by
a constant phase while keeping all the   other fields constant.
However, the difference will be relevant in a non-contractible
topology. This feature has been overlooked in the past. Indeed, in
the physical literature the gauge principle only appears in its
strong version.

The gauge principle has been generalized to non-Abelian groups and
has found extensive applications in quantum field theory. In the
Standard Model the electromagnetic gauge transformation is
imbedded in a non-trivial way in the electroweak gauge group
$SU(2) \otimes U(1)$. In this case we can think of the $U(1)$
gauge invariance of the present work as the $U(1)$ sector of the
electroweak group. Indeed, the quantization of the electric charge
follows from the quantization of the $U(1)$ sector (its charge is
usually denoted with the letter $Y$). For simplicity, but without
loss of generality, we shall ignore this aspect here. We shall
refer to the $U(1)$ gauge invariance as the electromagnetic gauge
invariance and to the constants $q_{\psi}$, $q_{\phi}$, as the
electric charges.

We also stress that in the whole work we shall remain in  a
classical field theory approach and we will never be involved with
the quantum theory. More precisely, the only quantum feature that
we shall use is that of considering a particle as mathematically
represented by a wave function rather than by a worldline; the
dynamics, however, will remain completely classical, i.e. dictated
by a Lagrangian and we will never use second quantization
procedures. Moreover we shall never be involved with explicit
expressions for the action or the observables. Our assumption is
that the action and the observables have values dependent only on
the physical state and not on its representant. Therefore they
should be {\em gauge invariant}.

\section{The gauge principle in a non-trivial
topology}\label{gau-prin-ntt}

Let $M$ be a curved spacetime with a metric $g_{\mu \nu}$ of
signature $(+ - - - )$. Let $F$ be a closed 2-form field on $M$
which we do not assume to be contractible. Let $\{ U_{i} \}$ be a
good covering of $M$ i.e. the open sets $U_{\alpha}$ are
contractible and so are their arbitrary intersections \cite[p.
28]{brylinski93}. We denote by $U_{i j}=U_{i} \cap U_{j}$, $U_{i j
k}=U_{i} \cap U_{j} \cap U_{k}$, the double and triple
intersections, respectively. Note that the Latin index refers to
the open set considered while the spacetime tensor indices are
denoted with Greek letters. We have no compelling experimental
evidence that our spacetime is not contractible thus we have in
fact no evidence that more than one open set are required to cover
$M$. Nevertheless, new investigations on the topology of the
Universe had appeared in recent years \cite{lachiezerey95,levin98}
and even some evidence of a non-trivial cosmological topology has
been pointed out \cite{luminet03}. The topological argument
explores the consequences of a non-contractible topology
assumption. We shall give here a form to the original argument
given by Dirac that is clearly related to the concept of \v{C}ech
cohomology. In fact \v{C}ech cohomology studies if a principal
$U(1)$ fibre bundle can be constructed over $M$ and the Dirac
argument essentially implies that the conditions that allow its
existence are fulfilled.

In the previous section we have introduced the gauge principle. It
is natural to generalize it to the present situation where we have
many open sets $U_{i}$. First, we can repeat the same steps as
before and see that in each open set $U_i$, given $F$ on $M$, we
have a potential $A^{i}$. Thus we can define what is a gauge
transformation in each $U_{i}$. The same holds for a generic
contractible open set $U \subset M$ for which a potential $A^U$
can be defined. In general if the set considered belongs to the
good covering we denote the superscript with $i$ in place of $U$.
Here the contractible open sets play the role of the contractible
spacetime considered previously. Note that there are special
fields that are directly observable like $F$ and $g$ while others
are not, like the electromagnetic fields $A^{i}$ that receive for
this reason an index of the open set where they are defined. The
reason is that in principle we may allow some discontinuity
between $A^{i}$ and $A^{j}$ in $U_{i j}$ as only the continuity of
observable quantities really matters.


It remains the question of the observability of matter fields. If
they are directly observable then we should not add to them an
index corresponding to the open set. However, we want to reproduce
the previously stated gauge principle in a given open set $U$ and
from that we already know that $\psi$ can not be observed
completely because of the gauge principle. Thus we add an index
$i$, and write $\psi^{i}$ in correspondence of the open set
$U_{i}$. In general the field will have a representant $\psi^U$ in
each contractible open set $U$ and we shall regard the triplet
$(A, \psi, \phi)$ as the collection $\{ (A^{U}, \psi^{U},
\phi^{U}) \}$ for $U \subset M$ contractible. Given the matter
fields and the potential on each open set $U$, we shall use the
notation $(A,\psi,\phi)\equiv \{(A^U,\psi^U,\phi^U)\}$.

We define a { weak \em gauge transformation on} $U_{i}$ as
\begin{eqnarray}
{A'}^{i}&=&A^{i}+\dd \alpha^{i}, \label{gauge1}\\
{\psi'}^{i}&=& e^{i q_{\psi} \alpha^{i}} \psi^{i} ,\label{gauge2}\\
{\phi'}^{i}&=&e^{i q_{\phi} \bar{\alpha}^{i}} \phi^{i}
,\label{gauge3}
\end{eqnarray}
where $h^i=\alpha^i-\bar{\alpha}^i$ is a constant, and analogously
for more general contractible open sets $U$ not belonging to
$\{U_i\}$. The strong gauge transformation satisfies $h^i=0$. We
shall also say that the fields $(A',\psi',\phi')$ and
$(A,\psi,\phi)$ are gauge related in $U$.
Then the {\em (weak/strong) gauge principle} is generalized as\\

{\em {\bf (Weak/strong) Gauge principle}. The configurations are
those collections $\{ (A^{U}, \psi^{U}, \phi^{U}) \}$ such that if
$U \subset V$, $(A^{V}, \psi^{V}, \phi^{V})\vert_{U} \sim (A^{U},
\psi^{U}, \phi^{U})$ where $\sim$ means that the fields between
brackets are gauge related.

The physical states of the theory are the equivalence classes
$[A,\psi,\phi]$ of configurations, where two configurations $(A',
\psi',\phi')$, $( A, \psi,\phi)$ are equivalent, $(A',
\psi',\phi')\sim( A, \psi,\phi)$,  iff for each open set $U$ they
are related by a gauge transformation for a
suitable function $\alpha^U$. }\\

The set of physical states will be denoted by $S=\{
[A,\psi,\phi]\}$.

The specification of a configuration $(A^i, \psi^i,\phi^i)$ in the
good covering $\{U_i \}$ determines uniquely the state since it
determines a representant in each contractible open subset.

\begin{remark} {\em
If the charges are quantized $q_{\psi}=m q$, $q_{\phi}=n q$ with
$m$ and $n$ coprime then, since what really matters is the phase
factor in the gauge transformation, the physics is determined by
the class whose elements are related by ${\alpha'}^{i}=\alpha^{i}+
a^{i}$ with $a^{i} \in \frac{2\pi}{m q} \mathbb{Z}$ (and
analogously for $\bar{\alpha}^{i}$ and $\bar{a}^{i}$). In the
strong case since $\alpha^{i}=\bar{\alpha}^{i}$,
${a}^{i}=\bar{a}^{i}$, in the end we have  $a^{i} \in
\frac{2\pi}{q} \mathbb{Z}$. In the weak case $\alpha$ and
$\bar{\alpha}$ can be redefined freely as above and therefore
$h^{i} \in \mathbb{R}/(\frac{2\pi}{mnq} \mathbb{Z})$.

If the charges are not quantized and we are in the strong case
$\alpha^{i} \in \mathbb{R}$, $a^{i}=\bar{a}^i=0$.

If  we are in the weak case, regardless of the quantization of
charges, $\alpha^{i} \in
\mathbb{R}/(\frac{2\pi}{q_{\psi}}\mathbb{Z})$, $a^{i} \in
\frac{2\pi}{q_{\psi}}\mathbb{Z}$, and analogously for
$\bar{\alpha}^{i}$, $\bar{a}^i$. Denoting by $\Lambda$ the
additive group given by the finite linear integer combinations of
the real numbers $2\pi q_{\psi}^{-1}$, $2 \pi q_{\phi}^{-1}$, we
have $h^{i} \in \mathbb{R}/\Lambda$. The quantized case considered
previously is therefore a particular case of this one.

 }
\end{remark}

%

\subsection{Intersection rule}
The gauge principle implies that for a given configuration if $U
,V \subset M$, $(A^{V}, \psi^{V}, \phi^{V})\vert_{U\cap V} \sim
(A^{U}, \psi^{U}, \phi^{U})\vert_{U\cap V}$. Let $U=U_{i}$ and
$V=U_{j}$ then $U \cap V=U_{i j }$ and we find that there are
functions $\beta^{ij} \, (\bar{\beta}^{ij}): U_{i j} \to
\mathbb{R}$ such that
\begin{eqnarray}
A^{i}&=&A^{j}+\dd \beta^{i j} , \label{int1}\\
\psi^{i}&=& e^{i q_{\psi} \beta^{i j}} \psi^{j} , \label{int2} \\
\phi^{i}&=&e^{i q_{\phi} \bar{\beta}^{i j}} \phi^{j} ,\label{int3}
\end{eqnarray}
where $k^{ij}=\beta^{i j}-\bar{\beta}^{i j}$ is a constant which
vanishes in the strong case. The first equation follows
necessarily from the fact that both potentials have the same
exterior differential in $U_{i j}$; the remaining two require the
gauge principle. We shall refer to the system
(\ref{int1})-(\ref{int3}) as the {\em intersection rule} since it
takes place in the intersection $U_{i j}$ and relates fields from
different sets.

Without loss of generality we can assume $\beta^{ij} = -
\beta^{ji}$ ($\bar{\beta}^{ij} = - \bar{\beta}^{ji}$). Indeed Eq.
(\ref{int1}) for the pairs (i,j) and (j,i) implies that
$\beta^{ji}$ and $-\beta^{ij}$ differ by a constant and thus we
can use the latter instead of the former in Eq. (\ref{int1}) for
the pair (j,i). Analogously Eq. (\ref{int2}) for the pairs (i,j)
and (j,i) implies that $e^{i q_{\psi} \beta^{j i}}= e^{-i q_{\psi}
\beta^{i j}}$, thus again $-\beta^{ij}$ can be used in place of
$\beta^{ji}$ and analogously for $\phi$.

\begin{remark} \label{kjh}{\em
If the charges are quantized $q_{\psi}=m q$, $q_{\phi}=n q$ with
$m$ and $n$ coprime then, since what really matters is the phase
factor in the intersection rule,  the physics is determined by the
class whose elements are related by ${\beta'}^{ij}=\beta^{ij}+
o^{ij}$ with $o^{ij} \in \frac{2\pi}{m q} \mathbb{Z}$ (and
analogously for $\bar{\beta}^{ij}$ and $\bar{o}^{ij}$). In the
strong case since $\beta^{ij}=\bar{\beta}^{ij}$,
${o}^{ij}=\bar{o}^{ij}$ we have $o^{ij} \in \frac{2\pi}{q}
\mathbb{Z}$. In the weak case $\beta$ and $\bar{\beta}$ can be
redefined freely as above and therefore $k^{ij} \in
\mathbb{R}/(\frac{2\pi}{mnq} \mathbb{Z})$.

If the charges are not quantized and we are in the strong case
then $\beta^{ij} \in \mathbb{R}$,  and $o^{ij}=\bar{o}^{ij}=0$.

If   we are in the weak case, regardless of the quantization of
charges, $\beta^{ij} \in
\mathbb{R}/(\frac{2\pi}{q_{\psi}}\mathbb{Z})$, $o^{ij} \in
\frac{2\pi}{q_{\psi}}\mathbb{Z}$, and analogously for
$\bar{\beta}^{ij}$, $\bar{o}^{ij}$. Moreover,  $ k^{ij} \in
\mathbb{R}/\Lambda$. The quantized case considered previously is
therefore a particular case of this one.

 }
\end{remark}

Assume that the physical theory considered satisfies  the gauge
principle and hence the intersection rule. The actual physical
problem depends both on $F$ and on the functions $\beta^{ij}$ (and
$\bar{\beta}^{ij}$) that appear in the intersection rule. Indeed
$\beta^{ij}$ does not depend solely on $F$ and on the gauge choice
of $A^{i}$ in each set. Let us comment this more extensively.
First note that $\beta^{i j}$ changes under a gauge transformation
on each $U_{i}$, ${A'}^{i}=A^{i}+ \dd \alpha^{i}$ as
\begin{equation} \label{bah}
{\beta'}^{i j}=\beta^{ij}+\alpha^{i}-\alpha^{j}.
\end{equation}
Now fix the gauge, i.e. a choice of $A^{i}$ in each set. This
choice fixes $\beta^{i j}$ only up to an additive constant, that
is, the knowledge of $F$ and the assumption that the intersection
rule (\ref{int1}) holds do not determine $\beta^{i j}$ completely.
Thus the physical problem is determined both by $F$ and, given a
choice of gauge in each set, by a $\beta^{i j}$ compatible with
that choice (i.e satisfying (\ref{int1})). We shall call that
$\beta^{i j}$ {\em the physical} $\beta^{i j}$ since suitable
physical experiments may measure its value up to transformations
${\beta'}^{ij}=\beta^{ij}+ o^{ij}$ with $o^{ij}$ ranging in a
suitable domain (Remark \ref{kjh}) for a given gauge choice.

This is exactly what happens in the Aharonov-Bohm effect. Let
$\gamma$ be a closed curve on spacetime. Choose events $e$ on the
curve and on suitable intersections $U_{ij}$ in  a way such that the
curve $\gamma$ between two successive events along the curve, lies
entirely in the open set $U_{i}$. For convenience relabel the sets
in such a way that $i$ takes the successive values $i=1,\ldots, N$
and identify $N+1$ with $1$.  Denote by $e_i$ the successive events,
$e_{i} \in U_{i-1 \, i}$.  The interference of a matter field with
itself in an Aharonov-Bohm experiment along the closed curve
$\gamma$ is determined by the gauge invariant quantity which does
not depend on the choice of $\{e_i\}$ \cite{wu75,alvarez85}
\begin{equation} \label{ph1}
\Phi[\gamma]=\exp\{ iq_{\psi}\sum_{i}[ ( \int_{\gamma
\,e_{i}}^{e_{i+1}} \!\!\! \! \!\!\! A^{i})+\beta^{i \, i-1}(e_i)]
\} .
\end{equation}
Clearly the functions $\beta^{ij}$ are not defined up to an
arbitrary constant, otherwise the interference would be arbitrary.
Indeed, they are defined, in a given gauge, only up to terms
$\frac{2\pi}{q_{\psi}}\mathbb{Z}$. Therefore there is some more
physical information encoded in the functions $\beta^{ij}$ than
what is determined alone by  Eq. (\ref{int1}). If $\gamma$ is a
boundary,  $\gamma=\p \Gamma$, it is not difficult to show that
\begin{equation}
\sum_{i}[  (\int_{\gamma \,e_{i}}^{e_{i+1}} \!\!\! \! \!\!\!
A^{i})+\beta^{i\, i-1}(e_i)]=\int_{\Gamma} F-\!\!\!\!\sum_{U_{ijk}:
\Gamma \cap U_{ijk}\ne 0 } c_{ijk}
\end{equation}
where in the next section
$c_{ijk}=\beta^{ij}+\beta^{jk}+\beta^{ki}$ will be proved to be
constant coefficients belonging to
$\frac{2\pi}{q_{\psi}}\mathbb{Z}$. As a consequence if $\gamma$ is
contractible to a point
\begin{equation} \label{ph2}
\Phi[\gamma]=\exp\{ iq_{\psi}\int_{\Gamma}F \} .
\end{equation}
which is the better known, but less general, expression for the
interference phase. In a contractible topology the weak gauge
principle would therefore lead to the usual phenomenology, i.e.
the interference phase measured in the Aharonov-Bohm experiment
would have the usual expression. However, in a truly
non-contractible spacetime there does not exist an a priori
constraint for the physical $\beta^{ij}$  and therefore $k^{ij}$
may differ from zero leading to a topological Aharonov-Bohm
interference even for neutral particles (see Sect. \ref{inte}).
This observation clarifies that strong and weak gauge principles
can be distinguished in a spacetime having non-contractible
topology. Since at present, the topology of the Universe has not
yet be determined, we can not yet rule out the weak gauge
principle possibility.

Summarizing we can  say that the physics is determined by a class
$[A^{i}, \beta^{ij} (, \bar{\beta}^{ij}), \psi^{i}, \phi^{i}]$
(each term is regarded as a set of maps, for instance $\beta^{ij}$
represents the set of maps $\beta^{ij}: U_{ij} \to \mathbb{R}$)
satisfying the intersection rule, and where two elements in the
class are related by
\begin{eqnarray*}
\!\!\!\!\!\!\!\!\!\!\!\!\!\!\!\!\!\!\!\!\!\!\!\!\!\!\!\!({A'}^{i},
{\beta'}^{ij} (,{\bar{\beta}'} {}^{ij}), {\psi'}^{i}, {\phi'}^{i})
\sim (A^{i}+\dd \alpha^{i}, \beta^{ij}+\alpha^{i}-\alpha^{j}
  (,\bar{\beta}^{ij}&+&\bar{\alpha}^{i}-\bar{\alpha}^{j}), e^{i
q_{\psi} \alpha^{i}} \psi^{i}, e^{i q_{\phi} \bar{\alpha}^{i}}
\phi^{i})
\end{eqnarray*}
with $\alpha^{i}$ suitable functions (note that
${k'}^{ij}=k^{ij}+h^i-h^j$).

\section{The topological argument}
Let us come to the core of the topological argument. We consider
here only the field $\psi$ but analogous considerations hold for
$\phi$. The reader will recognize the relation with \v{C}ech
cohomology.

Consider a triple intersection $U_{i j k}$. We already know that
in that set there is a function $\beta^{ij}$ such that
$A^{i}=A^{j}+\dd \beta^{i j}$ and analogously for the pairs
$(i,k)$, $(j,k)$. Moreover, summing up the three equations just
obtained we get
\begin{eqnarray} \label{coc}
&&\dd (\beta^{ij}+\beta^{jk}+\beta^{ki})=0 \quad \Rightarrow \nonumber \\
 &&c_{ijk}=\beta^{ij}+\beta^{jk}+\beta^{ki}=\textrm{cnst.} \quad
\textrm{on } U_{ijk}.
\end{eqnarray}
The reader familiar with \v{C}ech cohomology may realize that the
constants $c_{ijk}$ define a class in the \v{C}ech cohomology
group $H^{2}(M,\mathbb{R})$, two elements in the same class being
related by $c'_{ijk}= c_{ijk}+o^{ij}+o^{jk}-o^{ik}$ for suitable
constants $o^{ij}$. In fact, there is an isomorphism between  the
de Rham cohomology group $H^{2}_{dR}(M,\mathbb{R})$, to which
$[F]$ belongs, and the \v{C}ech cohomology group
$H^{2}(M,\mathbb{R})$ that sends $[F]$ to $[c_{ijk}]$. However,
note that we are identifying a precise representant of
$[c_{ijk}]$, the {\em physical $c_{ijk}$}, thanks to the
information that comes from the physical $\beta^{i j}$ i.e. the
one that satisfies all the intersection equations and not only
(\ref{int1}). Repeating the same calculations for $\phi$ we find
that
\begin{equation} \label{cbarc}
c_{ijk}-\bar{c}_{ijk}=k^{ij}+k^{jk}+k^{ki} ,
\end{equation}
and therefore $[c_{ijk}]=[\bar{c}_{ijk}]$ as classes  belonging to
$H^{2}(M,\mathbb{R})$ even in the weak gauge principle case
although the actual representant may change from matter field to
matter field.

\begin{remark} \label{r1} {\em
Consider the special case in which the charges are quantized. The
physical $\beta^{i j}$ is itself undetermined. Thus changing
${\beta'}^{ij}=\beta^{ij}+ o^{ij}$ we obtain
$c'_{ijk}=c_{ijk}+o^{ij}+o^{jk}+o^{ki}$.  In the strong case $o^{ij}
\in (2\pi/q) \mathbb{Z}$ thus the intersection rules (\ref{int1}),
(\ref{int2}) and (\ref{int3}) determine $[q c_{ijk}] \in
Z^{2}(M,\mathbb{R})/B^{2}(M,{2\pi}\mathbb{Z})$ while $[F]$ alone
only determines  $[c_{ijk}] \in H^{2}(M,\mathbb{R})$. In the weak
case $o^{ij} \in (2\pi/m q) \mathbb{Z}$,  $\bar{o}^{ij} \in (2\pi/n
q) \mathbb{Z}$ thus $[mqc_{ijk}] \in
Z^{2}(M,\mathbb{R})/B^{2}(M,{2\pi}\mathbb{Z})$ and analogously for
$[nq\bar{c}_{ijk}]$.  Moreover, as elements of
$Z^{2}(M,\mathbb{R})/B^{2}(M,{2\pi}\mathbb{Z})$ the classes $n[m q
c_{ijk}]$ and  $m[n q \bar{c}_{ijk}]$ are in general different.

If the charges are not quantized and we are in the strong case
$[c_{ijk}], [\bar{c}_{ijk}] \in Z^{2}(M,\mathbb{R})$.

If the charges are not quantized and we are in the weak case
$[q_{\psi}c_{ijk}] \in Z^{2}(M,\mathbb{R})/B^{2}(M,2 \pi
\mathbb{Z})$ and analogously for $[q_{\phi}\bar{c}_{ijk}]$.

 }
\end{remark}

The topological argument goes on to prove that $c_{ijk}$ are integer
constants up to a common factor.  Indeed, let us use repeatedly the
intersection rule for $\psi$ on $U_{ijk}$
\begin{eqnarray} \label{rer}
\psi^{i}&=& e^{i q_{\psi} \beta^{i j}} \psi^{j}= e^{i q_{\psi}
(\beta^{i j}+\beta^{j k})} \psi^{k} \nonumber \\
&=& e^{i q_{\psi} (\beta^{i j}+\beta^{j k}+\beta^{k i})}
\psi^{i}=e^{i q_{\psi} c_{ijk}} \psi^{i}.
\end{eqnarray}
Since this equation holds for any field $\psi$ we have
\begin{equation} \label{f1}
q_{\psi} c_{ijk}=2\pi m_{ijk} ,
\end{equation}
with $m_{ijk} \in \mathbb{Z}$. Repeating the argument for $\phi$,
\begin{equation} \label{f2}
q_{\phi} \bar{c}_{ijk}=2\pi n_{ijk} ,
\end{equation}
with $n_{ijk} \in \mathbb{Z}$, and using the relation between
$c_{ijk}$ and $\bar{c}_{ijk}$ we obtain (we assume the charges to
be different from zero, otherwise the issue of the quantization of
charge would have a trivial affirmative answer)
\begin{equation} \label{dsf}
\frac{2\pi}{q_{\psi}}
m_{ijk}-\frac{2\pi}{q_{\phi}}n_{ijk}=k^{ij}+k^{jk}+k^{ki} .
\end{equation}

We are going to separate the study into the strong and the weak
cases. Although, the former will turn out to be a special case of
the latter we shall consider them separately.

\subsection{The strong case}
In the strong case ($k^{ij}=0$) the previous  equation implies that
if for a certain $U_{ijk}$, $\bar{c}_{ijk} \ne 0$ (and hence
$n_{ijk} \ne 0$) then $q_{\psi}/q_{\phi}$ is rational and therefore
the charges are quantized. In this case there are $m$, $n$ coprime
integers such that $q_{\psi}=m q$, $q_{\phi}=n q$
\begin{equation}
\frac{q_{\psi}}{q_{\phi}}=\frac{m_{ijk}}{n_{ijk}}=\frac{m}{n} ,
\end{equation}
thus $m_{ijk}/m \in \mathbb{Z}$, $n_{ijk}/n \in \mathbb{Z}$ and
hence $q c_{ijk} \in {2\pi}\mathbb{Z}$. Moreover, since the
charges are quantized the physical $\beta^{ij}$ is determined only
up to transformations ${\beta'}^{ij}=\beta^{ij}+o^{ij}$ with
$o^{ij} \in \frac{2\pi}{q}\mathbb{Z}$ and hence $[ q c_{ijk}] \in
H^{2}(M,{2\pi}\mathbb{Z})$ or $[\frac{qF}{2\pi}] \in
H^{2}(M,\mathbb{Z})$.
 If  for any $U_{ijk}$, $c_{ijk} = 0$ one can not infer whether
the charges are quantized or not, and in the former case, by the
same argument used above, $[q c_{ijk}]\in
H^{2}(M,{2\pi}\mathbb{Z})$ is the trivial class.

Thus we have proved that the are only two possibilities:
\begin{itemize}
\item[(A1)] The charges are not quantized and
 for any $U_{ijk}$ the physical $c_{ijk}$ satisfies $c_{ijk}=0$.
\item[(A2)] The charges are quantized and $[q c_{ijk}] \in
H^{2}(M,{2\pi}\mathbb{Z})$ (i.e. $[\frac{q}{2\pi}F] \in
H^{2}(M,\mathbb{Z})$).
\end{itemize}
These two possibilities express the old Dirac's quantization
argument in the language of \v{C}ech cohomology. Note that in both
cases there exists a principal $U(1)$-bundle with transition
functions $g_{ij}=e^{iq \beta^{ij}}$,  since $q c_{ijk} \in {2
\pi}\mathbb{Z}$. The matter fields can then be regarded as
sections of bundles  associated to $Q$ through representations $
\rho : U(1) \to GL(1,\mathbb{C})$, where $\rho$ is $u \to u^{m}$
for $\psi$ and $u \to u^n$ for $\phi$.

If $F$ is exact this mathematical setting for the matter fields is
not compulsory. Indeed, in that case there are constants $b^{ij} \in
\mathbb{R}$ such that
\begin{equation}
(\beta^{ij}-b^{ij})+(\beta^{jk}-b^{jk})+(\beta^{ki}-b^{ki})=0 .
\end{equation}
Therefore, there exists a $(\mathbb{R},+)$-principal bundle $R$
with transition functions $g_{ij}=\beta^{ij}-b^{ij}$. The matter
fields can then be regarded as sections of vector bundles
associated to $R$.

\subsection{The weak case} \label{5b}
In the weak case the deduction of the quantization condition is
more involved. We said that $[c_{ijk}]=[\bar{c}_{ijk}]$ as classes
belonging to $H^{2}(M,\mathbb{R})$. This should be expected since
the isomorphism between $H^{2}(M,\mathbb{R})$ and
$H^{2}_{dR}(M,\mathbb{R})$ associates to both $[c_{ijk}]$,
$[\bar{c}_{ijk}]$ the class $[F]$ of the electromagnetic field. By
Remark \ref{kjh} and Eq. (\ref{dsf}) it follows that $[k^{ij}] \in
H^{1}(M, \mathbb{R}/\Lambda)$. We mention that  $H^{1}(M,
\mathbb{R}/\Lambda)$ is the set of isomorphism classes of
principal $\mathbb{R}/\Lambda$-bundles. For a treatment of
$\mathbb{R}/\Lambda$-bundles we refer the reader to \cite[Sect.
2.5]{brylinski93}.

There are two cases

{\bf 1}).  If the class $[c_{ijk}]=[\bar{c}_{ijk}] \in
H^{2}(M,\mathbb{R})$ is not trivial ($F$ is not exact) by
Poincar\'e duality we can find a closed surface $S$ such that
$\int_{S} F \ne 0$. It can also be shown that this integral can be
expressed as the sum of some coefficients $c_{ijk}$ such that
$U_{ijk} \cap S \ne 0$ (see for instance \cite{alvarez85}) and
analogously for $\bar{c}_{ijk}$. The equations (\ref{f1}) and
(\ref{f2}) imply
\begin{eqnarray}
q_{\psi} \int_{S}F &\in &  2\pi \mathbb{Z}, \\
q_{\phi} \int_{S}F &\in &  2\pi \mathbb{Z},
\end{eqnarray}
and therefore the charges are quantized. Substituting $q_{\psi}=m
q$, $q_{\phi}=n q$ and using the fact that $m$ and $n$ are coprime
we obtain $\int_{S}F \in \frac{2\pi}{q} \mathbb{Z}$, that is
$[\frac{q}{2\pi}F] \in H^{2}(M, \mathbb{Z})$. From Eq. (\ref{f2})
and (\ref{cbarc})
\begin{equation}
n c_{ijk}=\frac{2 \pi}{q} n_{ijk}+n (k^{ij}+k^{jk}+k^{ki}) .
\end{equation}
Using Eq. (\ref{f1}) we obtain ($n N+m M=1$)
\begin{eqnarray}
\!\!c_{ijk}\!\!\!&=&\!\!\!\frac{2\pi}{q}(N n_{ijk}\!+\!M
m_{ijk})\!+\! N n
(k^{ij}\!+\!k^{jk}\!+\!k^{ki}) ,  \label{c1}\\
\!\!\bar{c}_{ijk}\!\!\!&=&\!\!\!\frac{2\pi}{q}(N n_{ijk}\!+\!M
m_{ijk})\!-\! M m (k^{ij}\!+\!k^{jk}\!+\!k^{ki}) . \label{c2}
\end{eqnarray}
Multiplying  Eq. (\ref{cbarc}) by $m n$ and using (\ref{f1}) and
(\ref{f2}) we obtain
\begin{equation} \label{eur}
mnq(k^{ij}+k^{jk}+k^{ki}) \in {2\pi} \mathbb{Z} ,
\end{equation}
therefore, as  (Remark \ref{kjh}) $mnq k^{ij} \in
\mathbb{R}/{2\pi} \mathbb{Z}$, the coefficients $mnq k^{ij}$
define a cocycle in the \v{C}ech cohomology.  It has been
previously pointed out that under gauge transformations
${k'}^{ij}=k^{ij}+h^i-h^j$, thus
\begin{equation} \label{bg}
[mnq k^{ij}] \in H^{1}(M, \mathbb{R}/(2\pi \mathbb{Z})) .
\end{equation}
In other words, in the weak quantized case there exists a flat
bundle $K$ with transition functions $e^{i  mnq k^{ij}}$. We call
the class $k=[mnq k^{ij}]$, the (relative) {\em interference
class}.

From Eq. (\ref{eur}) it follows that $m q c_{ijk} \in
2\pi\mathbb{Z}$ and analogously for $\bar{c}_{ijk}$. But we
already know  (Remark \ref{r1}) that $[m q c_{ijk}] \in
Z^{2}(M,\mathbb{R})/B^{2}(M,2\pi\mathbb{Z})$ and thus
\begin{eqnarray}
{}[m q c_{ijk}] &\in &  H^{2}(M,{2\pi}\mathbb{Z}), \label{nh}\\
{}[n q \bar{c}_{ijk}] &\in &  H^{2}(M,{2\pi}\mathbb{Z}) .
\end{eqnarray}
These classes can be trivial or not, in any case we can construct
two principal $U(1)$-bundles $P_{\psi}$ and $P_{\phi}$ with
transition functions $e^{imq\beta^{ij}}$ and $e^{i n
q\bar\beta^{ij}}$, respectively. Consider the short exact sequence
\begin{equation}
0\to {2\pi}\mathbb{Z}\to \mathbb{R}
\longrightarrow^{\!\!\!\!\!\!\!\!\!\!\!\!{ e^{ix}}} \   U(1)
\to 1 ,
\end{equation}
where we  identify $U(1)$ and $\mathbb{R}/{2\pi}\mathbb{Z}$. It
gives rise to the long exact sequence
\begin{eqnarray*}
&& \qquad \qquad 0 \to  H^{1}(M,2\pi\mathbb{Z}) \to H^{1}(M,\mathbb{R}) \longrightarrow^{\!\!\!\!\!\!\!\!\!\sigma} \\
&&  H^{1}(M,U(1)) \longrightarrow^{\!\!\!\!\!\!\!\!\!\eta} \ \
H^{2}(M,{2\pi}\mathbb{Z})\longrightarrow^{\!\!\!\!\!\!\!\!\!\gamma}
\ \ H^{2}(M,\mathbb{R}) \to \ldots ,
\end{eqnarray*}
where $\textrm{Im}(\eta)=\textrm{Ker}(\gamma)$ is the torsion
subgroup of $H^{2}(M,2\pi\mathbb{Z})$. The Eq. (\ref{cbarc})
multiplied by $mn$ reads
\begin{equation}
n[m q c_{ijk}]-m[n q \bar{c}_{ijk}]=\eta([mnq k^{ij}]) ,
\end{equation}
or
\begin{equation} \label{kk}
K= P_{\psi}^{n} \otimes P_{\phi}^{-m} .
\end{equation}
 If the manifold has
vanishing torsion then $K$ is trivial. One should be  careful here
because in general although
\begin{equation}
\eta([mnq k^{ij}])_{ijk}= mnq (k^{ij}+k^{jk}+k^{ki}) ,
\end{equation}
this class is not necessarily trivial as $mnq k^{ij}$ does not
take values in ${2\pi}\mathbb{Z}$.

Consider the functions
\begin{equation}
q^{ij}_{MN}=Mm\beta^{ij}+Nn \bar\beta^{ij},
\end{equation}
where the index $MN$ recall that the constants $M$ and $N$ such
that $Mm+Nn=1$ are not unique. If $M'$ and $N'$ is another pair
with the same property then it is easy to show that there exists
an integer $j$ such that $M'=M+jn$, $N'=N-jm$. Thus
\begin{equation}\label{qq}
q^{ij}_{M'N'}=q^{ij}_{MN}+j mn k^{ij}.
\end{equation}

The redefinition freedom of $\beta^{ij}$ and $\bar\beta^{ij}$
(Remark \ref{kjh}) implies that $q^{ij}_{MN}(x) \in
\mathbb{R}/\frac{2\pi}{q} \mathbb{Z}$. Moreover, according to Eqs
(\ref{c1}) and (\ref{c2})
\begin{eqnarray*}
\xi_{MN}^{ijk}&=&q^{ij}_{MN}+q^{jk}_{MN}+q^{ki}_{MN} \\ &=&
\frac{2\pi }{q}(N n_{ijk}\!+\!M m_{ijk}) \in \frac{2 \pi}{q}
\mathbb{Z},
\end{eqnarray*}
define a class $[q \xi_{MN}^{ijk}] \in H^{2}(M,2\pi\mathbb{Z})$
and the transition functions $e^{iq \,q^{ij}_{MN}}$ define a
principal $U(1)$-bundle $Q_{MN}$. The Eq. (\ref{qq}) implies that
\begin{equation}
Q_{M'N'}=Q_{MN} \otimes K^{j}.
\end{equation}
Since $K$ belongs to the torsion, $\gamma([q \xi_{MN}^{ijk}])$ does
not depend on the choice $M,N$. From Eq. (\ref{nh}) we have
\begin{equation}
\gamma([mq c_{ijk}])=m q [F] \in H^{2}(M,\mathbb{R}),
\end{equation}
and from Eqs. (\ref{bg}) and (\ref{c1})
\begin{equation}
\gamma([mq c_{ijk}])=m \gamma([q \xi_{MN}^{ijk}]) ,
\end{equation}
so that
\begin{equation}
\gamma([q \xi_{MN}^{ijk}])=q [F] \in H^{2}(M,\mathbb{R}).
\end{equation}
In other words, the classes  on $H^{2}(M,2\pi\mathbb{Z})$
associated to the principal bundles $Q_{MN}$ all project  on the
electromagnetic field class on $H^{2}(M,\mathbb{R})$. We find
again that $[\frac{qF}{2\pi}]$ is an integer class.

 The torsion subgroup is a finitely
generated Abelian group. The interference class determines a flat
bundle $K$ such that $K^{\text{ord} \eta(k)}$ is trivial, where
$\text{ord} \eta(k)$ is the order of the Abelian subgroup
generated by the image $\eta(k)$ of the interference class  on
$H^{2}(M, 2 \pi \mathbb{Z})$. As a consequence
\begin{equation}
Q_{M+n\, \text{ord}\eta(k) \,\, N-m\,  \text{ord}\eta(k) }=Q_{MN}
,
\end{equation}
and the principal bundle denoted by $Q^{\textrm{ord} \eta(k)}$ and
defined by
\begin{equation} \label{el}
Q^{\textrm{ord} \eta(k)}\equiv Q_{MN}^{\textrm{ord} \eta(k)}
\end{equation}
does not depend on the choice of $M,N$.

 Now,  note that
$m\beta^{ij}=m q^{ij}_{MN}+Nmnk^{ij}$ and $n\bar\beta^{ij}=n
q^{ij}_{MN}-Mmnk^{ij}$ so that
\begin{eqnarray}
P_{\psi}&=&Q_{MN}^{m}\otimes K^{N} \label{j1} \\
P_{\phi}&=&Q_{MN}^{n}\otimes K^{-M} \label{j2}
\end{eqnarray}
and
\begin{equation} \label{qmn}
Q_{MN}=P_{\psi}^{M}\otimes P_{\phi}^{N}
\end{equation}
which hold for every pair $M$,$N$, satisfying $Mm+Nn=1$.  If
$\eta(k)$ is trivial then $K$ is trivial,  $\text{ord}\eta(k)=1$
and $Q= Q_{MN}$ does not depend on the choice  $M$, $N$. Moreover,
the equations (\ref{j1})-(\ref{j2}) show that the bundles
$P_{\psi}$ and $P_{\phi}$ admit a well defined  root $Q$ as in the
strong case. Conversely, if a bundle $Q$ exist such that
$P_{\psi}=Q^{m}$ and $P_{\phi}=Q^{n}$ then Eq. (\ref{kk}) shows
that $K$ is trivial.


Assume that $K$ is trivial. We already know that there is a root
principal $U(1)$-bundle $Q$. Moreover, there are constants $K^{ij}
\in \frac{2\pi}{mnq}\mathbb{Z}$ such that
\[
k^{ij}+k^{jk}+k^{ki}=K^{ij}+K^{jk}+K^{ki} .
\]
Define $o^{ij}=-N n K^{ij} \in \frac{2\pi}{m q}\mathbb{Z}$ and
$\bar{o}^{ij}=M m K^{ij} \in \frac{2\pi}{n q}\mathbb{Z}$ so that
$K^{ij}=\bar{o}^{ij}-o^{ij}$. Redefine
$\beta'^{ij}=\beta^{ij}+o^{ij}$ and analogously for
$\bar{\beta}^{ij}$, then due to Remark \ref{r1}, Eq. (\ref{cbarc})
can be written
\begin{equation} \label{jhg}
c'_{ijk}=\bar{c}'_{ijk} ,
\end{equation}
or
\[
(\beta'^{ij}-\bar{\beta}'^{ij})+(\beta'^{jk}-\bar{\beta}'^{jk})+(\beta'^{ki}-\bar{\beta}'^{ki})=0
\]
 which can be regarded as the condition for the existence of a
$(\mathbb{R},+)$ principal bundle. Since the fiber is contractible
the bundle is trivial and hence there are functions
$\alpha^{i}(x)$ such that
$\beta'^{ij}=\bar{\beta}'^{ij}+\alpha^i-\alpha^j$ or
$k'^{ij}=\alpha^i-\alpha^j$. A weak gauge transformation that
sends $k'^{ij}$ to zero exists iff $\alpha^i=h^{i}+\alpha$ for a
suitable function $\alpha$ and for suitable constants $h^{i} \in
\mathbb{R}/ (\frac{2\pi}{mnq}\mathbb{Z})$. This condition is
satisfied iff $[mnq k'^{ij}] \in H^{1}(M,U(1))$ is trivial. Thus
if $[mnqk'^{ij}] \in H^{1}(M,U(1))$ is trivial we are actually in
the strong case as the condition $k^{ij}=0$ is preserved under
strong gauge transformations.


{\bf 2}). It could be that although  $[c_{ijk}] \in
H^{2}(M,\mathbb{R})$ is trivial (i.e. $[c_{ijk}] \in
B^{2}(M,\mathbb{R})$) the charges are not quantized. In this case
there are constants $b^{ij} \in \mathbb{R}$ such that
$c_{ijk}=b^{ij}+b^{jk}+b^{ki}$ thus
$(\beta^{ij}-b^{ij})+(\beta^{jk}-b^{jk})+(\beta^{ki}-b^{ki})=0$. The
functions $\beta^{ij}-b^{ij}$ can be regarded as the transition
functions of a  principal bundle of structure group $(\mathbb{R},+)$
as the previous condition states that the cocycle condition for the
transition functions is satisfied. Since the fiber is contractible
the principal bundle is trivial and therefore there are functions
$\alpha^{i}$ such that $\beta^{ij}-b^{ij}+\alpha^{i}-\alpha^{j}=0$.
This last equation means that there is a particular gauge in each
$U_i$ such that the functions $\beta^{ij}$ become constant,
${\beta}^{ij}=b^{ij} \in \mathbb{R}/\frac{2\pi}{q_{\psi}}\mathbb{Z}$
(where we have used the indeterminacy of $\beta^{ij}$) and
therefore, since
 the coefficients $k^{ij}$ are constant,
$\bar{\beta}^{ij}=b^{ij}-k^{ij} \in
\mathbb{R}/\frac{2\pi}{q_{\phi}}\mathbb{Z}$ are constant too. The
equations (\ref{f1}), (\ref{f2}), provide further constraints
\begin{eqnarray}
q_{\psi} (b^{ij}+b^{jk}+b^{ki})&=&2\pi m_{ijk}, \label{g1}\\
q_{\phi} (\bar{b}^{ij}+\bar{b}^{jk}+\bar{b}^{ki})&=&2\pi n_{ijk} .
\end{eqnarray}
We can associate to the constants $b^{ij}$ (resp. $\bar{b}^{ij}$)
a flat bundle of transition functions $e^{i q_{\psi}b^{ij}}$
(resp. $e^{i q_{\phi}\bar{b}^{ij}}$) which can be trivial or not.
From Eq. (\ref{g1})  it follows that $[q_{\psi}c_{ijk}] \in
H^{2}(M,{2\pi}\mathbb{Z})$ and analogously for
$[q_{\phi}\bar{c}_{ijk}]$. The classes of
$H^{2}(M,{2\pi}\mathbb{Z})$ that are trivial when considered as
classes of $H^{2}(M,\mathbb{R})$ belong to the {\em torsion} of
$H^{2}(M,{2\pi}\mathbb{Z})$.

Note that if $[q_{\psi}{c}_{ijk}]$, $[q_{\phi}\bar{c}_{ijk}], \in
H^{2}(M, {2\pi}\mathbb{Z})$ are trivial then there are functions
$\gamma^i$, $\bar\gamma^i$ such that $b^{ij}-\gamma^i+\gamma^j \in
\frac{2\pi}{q_{\psi}}\mathbb{Z}$ and
$\bar{b}^{ij}-\bar\gamma^i+\bar\gamma^j \in
\frac{2\pi}{q_{\phi}}\mathbb{Z}$. These equations mean that the
functions $\beta^{ij}$,  $\bar\beta^{ij}$, can be redefined so that
there exist functions $\alpha^i$, $\bar{\alpha}^i$ such that
$\beta^{ij}+\alpha^i-\alpha^j=0$,
$\bar\beta^{ij}+\bar\alpha^i-\bar\alpha^j=0$. In particular
 it is
possible to find a weak gauge transformation such that
$e^{iq_{\psi}\beta^{ij}}=1$ or $e^{iq_{\phi}\bar\beta^{ij}}=1$ but
 these gauges do not necessarily coincide. They coincide iff
there exist constants $h^i\in \mathbb{R}/\Lambda$ such that
$k^{ij}=h^{i}-h^{j}$ since in this case the terms $h^i$ can be
removed with a weak gauge transformation. Since  $[k^{ij}] \in
H^{1}(M,\mathbb{R}/\Lambda)$ we conclude that the problem reduces
to the strong case if $[k^{ij}]$ is trivial.

Summarizing, there are  two possibilities:
\begin{itemize}
\item[(B1)]  The charges are not quantized, $[k^{ij}]
\in H^{1}(M,\mathbb{R}/\Lambda)$, $[c_{ijk}]=[\bar{c}_{ijk}] \in
H^{2}(M,\mathbb{R})$ is trivial and $[q_{\psi}c_{ijk}]$,
$[q_{\phi}\bar{c}_{ijk}] \in H^{2}(M,{2\pi}\mathbb{Z})$. If these
three classes are trivial  we are in the strong case (A1).
\item[(B2)] The charges are quantized and $[m q c_{ijk}],[n q \bar{c}_{ijk}] \in
H^{2}(M,{2\pi}\mathbb{Z})$ satisfy $\gamma([m q c_{ijk}])=m[qF]$,
$\gamma([n q \bar{c}_{ijk}])=n[qF]$ and $[\frac{qF}{2\pi}]$ is an
integer class. There are principal $U(1)$-bundles $P_{\psi}$ and
$P_{\phi}$ associated to the classes $[m q c_{ijk}]$, $[n q
\bar{c}_{ijk}]$, that satisfy $P_{\psi}^{n}\otimes
P_{\phi}^{-m}=K$ where $K$ is a flat bundle associated to the
class $\eta([mnq k^{ij}])$ where $[mnq k^{ij}] \in H^{1}(M,U(1))$
is the interference class. $K$ is trivial iff a root principal
$U(1)$-bundle $Q$ exists such that $P_{\psi}=Q^{m}$,
$P_{\phi}=Q^{n}$. For every $M,N$ such that $Mm+Nn=1$ the Eqs.
(\ref{j1}), (\ref{j2}) and (\ref{qmn}) hold. In particular the
principal bundle defined by (\ref{el}) does not depend on the
choice of $M,N$. If $[mnq k^{ij}]\in H^{1}(M,U(1))$ is trivial we
are in the strong case (A2).
\end{itemize}

From the above study we conclude that in both the strong and weak
cases if the electromagnetic field is not exact the charges are
quantized (cases A2 and B2). In any case, independently of the
exactness of the electromagnetic field the most interesting case
is (B2) as the charges are observationally quantized and (A2) is a
special case of it.

A relevant difference between the weak and strong cases is that in
the weak case there could be non quantized charges with $[q_{\psi}
c_{ijk}] \in H^{2}(M,{2\pi}\mathbb{Z})$ non-trivial and $[q_{\psi}
c_{ijk}] \in H^{2}(M,\mathbb{R})$ trivial. Such $[q_{\psi}
c_{ijk}]$ are non-trivial torsion classes and generate non-trivial
flat bundles. If the weak gauge principle holds it is no longer
true that a particle description through non-trivial bundles
implies the quantization of charges. Moreover, the existence of a
root principal bundle $Q$ can not be inferred in the weak case.
The description of matter fields as sections of vector bundles
associated to the same universal bundle then radically changes.
Each particle has its own principal bundle.

The long exact sequence for the quantized case implies
\begin{equation}
\ker \eta=H^{1}(X,\mathbb{R})/H^{1}(X,\mathbb{Z}).
\end{equation}
Note that in a simply connected spacetime $\ker \eta=0$ since
$H^{1}(X,\mathbb{R})\sim H^{1}_{dR}(X, \mathbb{R})=0$ as in a
simply connected manifold all the closed 1-forms are exact. This
can be also seen from the universal covering theorem which states
that $H^{1}(X,A)\simeq Hom(\pi_{1}(X),A)$, however one should be
careful since $H^{1}(X,A)=0$  does not mean that $\pi_{1}(X)=\{ e
\}$. Moreover, in a simply connected spacetime the torsion of
$H^{2}(M,2\pi\mathbb{Z})$ vanishes since it is the image of
$H^{1}(M, U(1))$ under $\eta$.  We conclude that the strong case
is equivalent to the weak case in simply connected manifolds.

\section{Interpretation} \label{inte}
We give an interpretation of (B2) which is the most interesting
case from the physical point of view.

The generic matter field, say $\psi$, may not be described as the
section of a vector bundle associated to $Q$ under the
representation $\rho_{\psi}: U(1) \to GL(1, \mathbb{C})$, $u \to
u^{m}$ since the root bundle $Q$ does not always exist. On the
contrary, each field has its own principal bundle, for instance
$\psi$ is a section of a vector bundle associated to $P_{\psi}$
under the trivial representation $u \to u$. In general we can
regard every field as a section of a vector bundle associated to a
$U(1)$-bundle under the trivial representation. In this way the
different particles are in one-to-one correspondence with the
$U(1)$ principal bundles. The possibility of describing the fields
as sections of vector bundles associated with {\em the same}
principal bundle under non-trivial representations arises only if
the different $U(1)$-bundles considered have a common root. Note
that on $P_{\psi}$ a connection can be defined that takes, in
suitable local coordinates the form, $\omega_{\psi}=i(\dd
\alpha^{i}-m q A^{i})$, so that covariant derivatives of matter
fields make sense. However, no universal principal bundle $Q$ with
a universal connection of the form $\omega=i(\dd \alpha^{i}-q
A^{i})$ as in usual (strong) gauge theory exists. In any case, on
the principal bundles $Q_{MN}$ a connection of that form can be
defined although the principal bundle associated to the generic
particle will  not be always of the form $Q_{MN}^{a}$ for suitable
$a$ and $M,N$.

Consider a particle obtained as a bound state of $z_1$ particles
$\psi$ and $z_2$  particles $\phi$. The number $z_1$ and $z_2$ are
integers and if say $z_2$, is negative then there are $\vert
z_2\vert$ antiparticles of $\phi$ in the bound state. The actual
forces responsible for the bound state may not be of
electromagnetic origin and are not important for our analysis. The
new bound state is described by the principal bundle
$P_{\psi}^{z_1}\otimes P_{\phi}^{z_2}$.

The Eqs. (\ref{qmn}) and (\ref{kk}) show that all the $U(1)$
principal bundles of the form
\begin{eqnarray}
P&=&Q_{MN}^{a} \otimes K^{b} \qquad a \in \mathbb{Z}, \ b \in
\mathbb{Z}_{/\textrm{ord} \eta(k)} \\
&=& P_{\psi}^{Ma+nb}\otimes P_{\phi}^{Na-mb}
\end{eqnarray}
can be generated from the physical building blocks $P_{\psi}$ and
$P_{\phi}$, however, no common root exists. Under the one-to-one
identification of $U(1)$ principal bundles and fields, the quantity
$aq$ represents the field charge. In particular there are
$\textrm{ord} \eta(k)$ neutral particles ($a=0$),
\begin{equation}
K^{0}, \ K^1, \ldots, K^{\textrm{ord} \eta(k)-1}
\end{equation}
 which can also
be regarded as different topological vacuum states. They form a
group $\langle K \rangle $ isomorphic to
$\mathbb{Z}_{/\textrm{ord}\eta(k)}$ under tensorial
multiplication. Under the same operation they act on
$H^{2}(M,2\pi\mathbb{Z})$ separating it into cyclic orbits of
$\textrm{ord} \eta(k)$ elements each. For instance a principal
bundle $P_{\psi}$ whose class is in $H^{2}(M,2\pi\mathbb{Z})$
belongs to the orbit
\begin{equation}
P_{\psi}, \ P_{\psi}  \otimes K, \ldots,  \ P_{\psi} \otimes
K^{\textrm{ord} \eta(k)-1}.
\end{equation}
The neutral content of a generic charged particle is not
univocally determined. Indeed,
\begin{equation}
Q_{MN}^{a} \otimes K^{b}=Q_{M'N'}^{a} \otimes K^{b-j\, a}.
\end{equation}
The interpretation becomes clear looking at Eqs. (\ref{qmn}) and
(\ref{kk}). The particle represented by the principal bundle $P$
may be regarded as containing $aM$ particles $\psi$, $aN$
particles $\phi$ and $b$ neutral particles $K$. However, the
particles $\psi$ and $\phi$ can change in number according to Eq.
(\ref{kk}) as they can annihilate to form neutral particles $K$.
The neutral particle content can not in general be determined.
However, if $a$ is a multiple of $\textrm{ord} \eta(k)$,
$a=\tilde{a} \textrm{ord} \eta(k)$, then the constant $b$ does not
depend on the  choice $M,N$
\begin{eqnarray}
P&=&(Q^{\textrm{ord} \eta(k)})^{\tilde{a}} \otimes K^{b} \qquad
\tilde{a} \in \mathbb{Z} .
\end{eqnarray}
Thus the particles having a charge multiple of $\textrm{ord}
\eta(k) q$ have a special role as they have a well defined neutral
particle content.

Given two fields $\psi$ and $\phi$, the class $k=[mnq k^{ij}] \in
H^{1}(M, U(1))$ that we termed the (relative) {\em interference
class} determines the different behavior of the fields under
Aharonov-Bohm  interference caused by the topology of the
Universe. In other words the topology of the Universe (i.e. its
`holes'), being non-trivial, may act in a way analogous to the
solenoid in the Aharonov-Bohm experiment. However, contrary to
what could be naively expected from this analogy the interference
phases of $\psi$ and $\phi$ are not of the form $u^{m}$, $u^{n}$
for a suitable $u\in U(1)$. The interference class determines the
different way in which these particles couple with the topology of
the Universe.

We can see this fact easily from the expression of the
Aharonov-Bohm phase for the neutral  particle $K$. Using Eq.
(\ref{ph1})
\begin{eqnarray}
\Phi_{k}[\gamma]&=&(\Phi_{\psi}[\gamma])^{n}(\Phi_{\phi}[\gamma])^{-m}=\exp\{
imnq\sum_{i} k^{i\, i-1} \} \nonumber \\
&=&\langle k , [\gamma]\rangle ,
\end{eqnarray}
where $\langle,\rangle$ is the dual pairing between $H^{1}(M,U(1))$
and $H_{1}(M,U(1))$, and $[\gamma]$ is the homology class whose
representant is $\gamma$.

A particular case is obtained if $\eta(k)$ is trivial in
$H^{2}(M,2\pi \mathbb{Z})$, i.e. if $K$ is trivial. In this case
$k=\sigma(k_{\mathbb{R}})$ for a suitable class $k_{\mathbb{R}}
\in H^{1}(M,\mathbb{R})$. Then
\begin{eqnarray}
\Phi_{k}[\gamma]&=& \langle k , [\gamma]\rangle =e^{i\langle
k_{\mathbb{R}} , [\gamma]_{\mathbb{R}}\rangle_{\mathbb{R}}} ,
\end{eqnarray}
where $\langle,\rangle_{\mathbb{R}}$ is the dual pairing between
$H^{1}(M,{\mathbb{R}})$ and $H_{1}(M,{\mathbb{R}})$, and
$[\gamma]_{\mathbb{R}}$ is the  corresponding homology class whose
representant is $\gamma$. In other words there is a closed 1-form
denoted again with $k_{\mathbb{R}}$ such that
$\Phi_{k}[\gamma]=\exp\{i\oint_{\gamma} k_{\mathbb{R}} \}$. Thus
there is a topological Aharonov-Bohm effect that acts on neutral
particles. It acquires its characteristic exponential form only if
the torsion of the particle vanishes.

\section{Quarks}

In the usual strong case there is only one charge quantization
unit $q$. Since the quarks are described in the Lagrangian by
matter fields, the quantization unit should be necessarily
identified with one third  the (minus) electric charge $q=e/3$ or
with a subunit $q/z$, $z\in \mathbb{Z}^{+}$. It is therefore
incorrect in the strong case to identify the quantum $q$ with
minus the electric charge $e$. However, it is an experimental fact
that all the observed particles have a charge that is a multiple
of $e=3q$ (quark confinement). The usual strong case does not
provide a mathematics sufficiently rich to describe such a
situation.

In the weak case we have seen that given two fields there are two
quantization units of relevance for the theory. The charge quantum
$q$ and the unit $\textrm{ord}\eta(k) q$. It is natural to
identify $e\equiv \textrm{ord}\eta(k) q$, and hence to assume that
$\eta(k)$ is a torsion class $K$ that generates a cyclic subgroup
of order three, $\textrm{ord}\eta(k)=3$. Next we assume for
simplicity that $H^{2}(M,2 \pi \mathbb{Z})$ has a torsion subgroup
which coincides with the cyclic subgroup $\langle K \rangle$.

As we have seen the cyclic group acts on the principal bundle of
the field and generates an entire orbit of $\textrm{ord}\eta(k)=3$
particles. We proceed with the following identification. The
fields considered are quarks and the elements of the orbits
correspond to the different colors of the same particle flavor.

For instance, let the two fields be the `up' and `down' quarks for a
certain color. They have charge $2q$ and $-q$ respectively. Then
there is a principal $U(1)$-bundle $Q$ such that the following
bundles are identified with those of $u(r)$, $u(g)$ and $u(b)$
\begin{equation}
u: \quad Q^2, \quad Q^2\otimes K, \quad  Q^2 \otimes K^2 .
\end{equation}
The actual identification of this bundles with the corresponding
color is important only up to cyclic permutations. Analogously the
bundles corresponding to $d(r)$, $d(g)$ and $d(b)$ are
\begin{equation}
d: \quad Q^{-1}, \quad Q^{-1}\otimes K, \quad  Q^{-1} \otimes K^2
.
\end{equation}
The next flavor generations $(c,s)$ and $(t,b)$ live on the same
principal bundles. For instance $c(r)$ is a section on the same
bundle of $u(r)$. Here, we are considering the simplest possible
model. There is enough room for many other possibility for
instance considering a torsion subgroup larger that those
considered here. To the antiparticles correspond the principal
bundles
\begin{eqnarray*}
{}\bar{u}: &{}& \quad Q^{-2}, \quad Q^{-2}\otimes K^{2}, \quad
Q^{-2} \otimes K, \\
{}\bar{d}: &{}& \quad Q, \quad Q\otimes K^{2}, \quad  Q \otimes K,
\end{eqnarray*}
and analogously for $(\bar{c},\bar{s})$ and $(\bar{t},\bar{b})$.

The $SU(3)$ color transformations are not ordinary matrices.
Indeed, in order to preserve the above correspondences under color
transformations we must generalize these matrices. In this model
the coefficients $B_{ij}$ of a color matrix $B$ are not
$\mathbb{C}$ numbers, instead they are sections of a complex
vector bundle associated to $K^{2(j-i)}$, that is, using the
identification between $U(1)$-bundles and fields
\begin{displaymath}
B \in \left(\begin{array}{ccc}
K^0 & K^2 & K \\
   K & K^0 & K^2 \\
   K^2 & K & K^0
\end{array}\right).
\end{displaymath}
The coefficients reduce to the usual  complex numbers only locally
for a given gauge choice.

With the above identifications, taking into account that every
baryon is a color singlet, we have that the principal bundle
associated to it has the form $(Q^{3})^k$ for a suitable integer
$k$. In particular, it has  charge $ke$. This means that the
$U(1)$ principal bundles associated to the  baryonic fields have
indeed a common root $Q^3$ which is natural to identify with the
principal bundle root of the leptonic fields. Therefore, in this
model, if quarks are not taken into account the usual (strong)
gauge theory  applies.

%
%

\section{Cosmological considerations}

In the previous sections we have shown that the quantization of
charge is implied by the non-exactness of the electromagnetic
field. This  result leads to a natural physical consequence: the
electromagnetic field may manifest its non-exactness at
cosmological scales where the non-trivial topology manifests
itself  and, moreover, it needs not to be singular as in the
Dirac's monopole example. However, the electromagnetic field can
be non-exact only if the spacetime manifold has a non-trivial
cohomology group $H^{2}_{dR}(M,\mathbb{R})$. We have shown that
in this case the electromagnetic field belongs actually to a non
trivial class of $H^{2}(M,\frac{2 \pi}{q}\mathbb{Z})$ where $q$ is
the charge quantum (this statement also holds in the weak gauge
principle case). Now, a globally hyperbolic spacetime is a product
$M= S \times \mathbb{R}$ where $t \in \mathbb{R}$ is a time
function \cite{bernal03}. If $x^{i}$, $i=1,2,3$ are coordinates on
$S$ the metric takes the form
\begin{equation}
\dd s^{2}=\chi^{2}(t,x) \dd t^{2}-R^{2}(t,x) h_{ij}\dd x^{i} \dd
x^{j} ,
\end{equation}
where $\chi$ and $R$ are suitable functions and $\textrm{det}h=1$.
The expansion scalar of the congruence of timelike curves  given
by $u=\frac{1}{\chi}\p_{t}$ is $\theta=3\p_{u}\ln R$, hence $R$
can be interpreted as the scale factor of the Universe. Since $M=S
\times \mathbb{R}$, the electromagnetic field class is
proportional to a non trivial cohomology class of $H^{2}(S,\frac{2
\pi}{q}\mathbb{Z})$. This means that the topological
non-trivialness of the electromagnetic field arises from its space
components i.e. from the magnetic components. Now, unless $[F] \in
H^{2}(S,\frac{2 \pi}{q}\mathbb{Z})$ is a torsion class
\cite{bott82} (this can not happen, otherwise $F$ would be exact,
see  Sect. \ref{5b})
 there is a surface $\Sigma$ in $S$  such that
 $c_{\Sigma}=\frac{q}{2\pi}\int_{\Sigma} [F]$ is a constant
different from zero known as the first Chern number relative to
$\Sigma$. The point is that since it is an integer this number can
not change as the Universe expands. Since the area of $\Sigma$
expands as $R^{2}$, if $f$ denotes the intensity of the
electromagnetic field as measured by a local inertial observer in
$\Sigma$, its value scales as $f \sim 1/R^{2}$. Thus the expansion
of the Universe implies that the actual value of the magnetic
field in the non-trivial class decreases. Using the same argument
we see that the local energy density of the field scales as
$1/R^{4}$ exactly as the incoherent radiation does. This kind of
behavior was present since the beginning of the Universe when its
topology acquired a final form (at least according to general
relativity). The possibility of measuring today a non-trivial
cosmological electromagnetic field is then almost ruled out. Even
if present, it would have now a negligible value due to the
expansion of the Universe. This conclusion is reinforced in those
cosmological scenarios that admit an initial inflation.

\section{Conclusions}

After introducing a weak gauge principle, which requires to consider
two charged fields, we have studied its implications for electric
charge quantization for generic spacetime topologies. We have shown
that this new gauge principle has nontrivial implications if the
spacetime has torsion in its second integral cohomology group, but
it coincides with the standard one if $H^1(M U(1))$ (and hence the
torsion) is trivial.

If the spacetime has torsion, we have shown that there exist
topologically non trivial configurations of charged fields which
do not imply charge quantization. This possibility has not been
previously recognized in the literature, although it is compatible
with present experimental knowledge. On the other hand, we have
proved that charges are quantized on any spacetime whenever the
electromagnetic field is not exact.

The weak gauge principle has been exploited, revealing on spaces
with torsion a richer structure than the one which would follow from
the ordinary gauge principle. In particular we have pointed out that
neutral particles can be affected by a topological Aharonov-Bohm
effect, and therefore the interference in such kind of experiments
does not depend solely on the charge of the particle considered but
on a topological invariant which is given by a flat line bundle. The
comparison of two different particles has revealed the role of the
interference class $k \in H^{1}(M, U(1))$. When this class is
trivial the phenomenology reduces to that of ordinary gauge theory.

We have shown that the weak gauge principle implies that the
$U(1)$-bundles determined by the matter fields are not, in general,
associated to a common principal bundle. Indeed, they are
expressible as a suitable power of a certain non-unique root bundle
times a torsion class. This torsion class plays the role of a new
quantum number. A torsion subgroup splits the space of
$U(1)$-bundles $H^2(M, \mathbb{Z})$ into orbits. Each orbit is
identified with a particle while its elements are identified with
the different quantum numbers of that particle. We have provided an
example considering the case of quarks, where the torsion subgroup
is the cyclic group $\mathbb{Z}/3$ and the quantum number generated
is the color of the particle. We have shown that the weak case is
more appropriate in order to describe the quarks and the particle
generated from them since the theory naturally embodies two
fundamental electric charges, the basic quark charge $q$ and the
electric charge $3q$.

We have also suggested that a weak non-exact component of the
electromagnetic field over cosmological scales could be
responsible for the quantization of the electric charge. Indeed,
due to the constancy of the Chern numbers characterizing the
non-trivial principle bundle, the expansion of the Universe would
make this non-trivial components negligible to the observational
capabilities of present day observers. In order to work, this
mechanism needs a non-contractible topology of the spacetime
manifold, although the non-trivial topology may manifest itself
only at cosmological scales. This interesting possibility has the
advantage of not being ruled out by observations although it has
the related disadvantage of being difficult to verify
experimentally.\\

\section*{Acknowledgments}
This work was developed during a two-year stay of E.M. at the
department of mathematics of Salamanca. E.M. would like to thank the
department of mathematics for kind hospitality and M. Castrillon for
a useful conversation on gauge theories. E.M.
is supported by INFN, grant No. 9503/02.\\


\end{document}